\documentclass[10pt,conference]{IEEEtran}

\usepackage{amsmath}
\usepackage{newtxtext,newtxmath}
\usepackage{graphicx}
\usepackage{textcomp}
\usepackage{xcolor}
\usepackage{booktabs}
\usepackage{tabularx}
\usepackage{url}
\usepackage{pgfplots}
\usepackage[numbers,sort&compress]{natbib}
\pgfplotsset{compat=1.18}
\usetikzlibrary{positioning,arrows.meta}
\definecolor{tealdark}{HTML}{1F6F78}
\definecolor{tealfill}{HTML}{E4F1F2}
\definecolor{indigodark}{HTML}{3B3F8C}
\definecolor{indigofill}{HTML}{E8E8F4}

\newcommand{\dsname}{PrivacyPR}
\newcommand{\dsdoi}{\url{https://doi.org/10.5281/zenodo.23744765}}

\usepackage{xcolor}
\usepackage[most]{tcolorbox}
\usepackage{listings}
\usepackage{tabularx}

\definecolor{exampleblue}{HTML}{356A9A}
\definecolor{examplebluebg}{HTML}{F2F7FB}
\definecolor{examplegreen}{HTML}{2E7D32}
\definecolor{examplegreenbg}{HTML}{F1F8F2}
\definecolor{examplegray}{HTML}{555555}
\definecolor{exampleborder}{HTML}{B8C8D8}

\newcommand{\field}[1]{%
  \textcolor{exampleblue}{\textbf{#1}}%
}

\lstdefinestyle{patchstyle}{
  basicstyle=\ttfamily\scriptsize,
  columns=fullflexible,
  keepspaces=true,
  showstringspaces=false,
  frame=none,
  aboveskip=2pt,
  belowskip=2pt,
  xleftmargin=3pt,
}

\begin{document}

\title{\dsname{}: A Dataset of Real-World Privacy Fixes for Benchmarking Privacy-Aware Code Review}

\author{\IEEEauthorblockN{Hua Yang\IEEEauthorrefmark{1},
Aditya Bhardwaj\IEEEauthorrefmark{1},
Shidong Pan\IEEEauthorrefmark{2} and
Bowen Xu\IEEEauthorrefmark{1}}
\IEEEauthorblockA{\IEEEauthorrefmark{1}North Carolina State University, Raleigh, NC, USA\\
\{hyang45, arbhardw, bxu22\}@ncsu.edu}
\IEEEauthorblockA{\IEEEauthorrefmark{2}Independent Researcher\\
Shidong.Pan@anu.edu.au}}

\maketitle

\begin{abstract}
Coding agents increasingly review code, yet whether they catch the privacy
defects that laws such as the GDPR require software to avoid remains largely untested, partly
because few datasets connect real privacy defects to their accepted fixes
and review context.
We present \dsname{}, a dataset of 4,131 merged GitHub pull requests (PRs) from
1,469 repositories, created between June 2018 and January 2024.
Each PR includes its commits, diffs, review discussion, and LLM-generated
labels for privacy relevance, fix type, and GDPR principles.
\dsname{} provides 1,169 high-confidence privacy fixes and 1,643 non-fix
PRs, together with \emph{Verified}, a manually checked split of 250 privacy
fixes and 250 non-fix PRs, and \emph{Small}, a 100 + 100 subset for
evaluation. On Verified, both annotators endorsed 72\% of the LLM's
GDPR-principle assignments.

Beyond supporting empirical study of how privacy defects arise and are
fixed, \dsname{} provides a benchmark for privacy-aware code review on
real, developer-accepted fixes. Its evaluation harness reverts each fix to
reintroduce the defect. On Small, two coding agents detect only 42--51\%
of defects under a generic review prompt, but 88--91\% when explicitly
asked to check privacy, while flagging 11--15\% of non-fix PRs.
These results show that privacy defects can be detected by coding agents,
but that detection depends strongly on review intent and comes with
false alarms.
Data, code, reviews, judge verdicts, and an extensible evaluation harness
are publicly available at \dsdoi{}.
\end{abstract}

\begin{IEEEkeywords}
privacy, GDPR, code review, pull requests, coding agents, dataset
\end{IEEEkeywords}

\section{Introduction}
\label{sec:intro}

Privacy laws such as the EU General Data Protection Regulation (GDPR)~\cite{gdpr} and the California Consumer Privacy Act (CCPA)~\cite{ccpa} require software to protect the personal data it processes.
Yet developers often find these requirements vague, costly, and difficult to operationalize~\cite{franke2024gdpr,li2020developers,tao2025privacy,pan2024trap}, and privacy violations often surface only after release, through audits of deployed applications~\cite{nguyen2021share} or reports in app reviews~\cite{nema2022analyzing} and issue trackers~\cite{hennig2026whos}.
Code review is therefore a critical opportunity to catch privacy defects before they are merged and shipped.
At the same time, code review is increasingly moving to coding agents.
Agents already resolve real GitHub issues (e.g.,~\cite{yang2024sweagent,zhang2024autocoderover}) and are deployed as reviewers: at Atlassian, 38.7\% of an LLM reviewer's comments led to code changes~\cite{tantithamthavorn2026rovodev}.
However, review agents solve only about 40\% of c-CRAB's tasks~\cite{zhang2026ccrab}, while existing automated-review benchmarks focus primarily on functional defects~\cite{zeng2025swrbench}, leaving non-functional requirements such as privacy comparatively unexplored.
As coding agents increasingly both produce and review code, an important question is whether they can identify privacy defects before those defects are merged.
Evaluating this capability requires real code changes in which a privacy defect can be linked to the fix developers accepted and the surrounding review context.
Such data are largely absent.

\begin{figure}[t]
\footnotesize
\begin{tcolorbox}[
  enhanced,
  colback=white,
  colframe=exampleborder,
  boxrule=0.6pt,
  arc=1.5mm,
  left=5pt,
  right=5pt,
  top=5pt,
  bottom=5pt,
  before skip=2pt,
  after skip=2pt
]

\begin{tcolorbox}[
  colback=examplebluebg,
  colframe=examplebluebg,
  boxrule=0pt,
  arc=1mm,
  left=4pt,
  right=4pt,
  top=3pt,
  bottom=3pt
]
{\color{exampleblue}\textbf{PR}}
\ \texttt{csharpfritz/CoreWiki \#278}
\hfill
{\color{examplegray}\scriptsize
437 stars $\cdot$ merged 2018-08-31 $\cdot$
1 commit $\cdot$ 1 file (+1/$-$0)}

\vspace{2pt}
\textbf{\emph{Prevent email notification being sent without user consent}}

{\color{examplegray}\scriptsize
Closes issue \#242 $\cdot$ keyword: \emph{user consent}}
\end{tcolorbox}

\vspace{3pt}

\field{Patch}
\quad
{\color{examplegray}\texttt{CoreWiki.Notifications/NotificationService.cs}}

\vspace{2pt}

\begin{tcolorbox}[
  colback=examplegreenbg,
  colframe=examplegreen!35,
  boxrule=0.4pt,
  arc=1mm,
  left=3pt,
  right=3pt,
  top=2pt,
  bottom=2pt
]
\begin{lstlisting}[style=patchstyle]
@@ -140,6 +140,7 @@ ... SendNewCommentEmail(...)
   if (!canNotifyUser())
   {
     _logger.LogInformation("... not consented ...");
+    return false;
   }
\end{lstlisting}
\end{tcolorbox}

\vspace{3pt}

\field{LLM annotation}

\vspace{2pt}

\begin{tabularx}{\linewidth}{@{}>{\bfseries\color{examplegray}}l X@{}}
Privacy related &
\textcolor{examplegreen}{\textbf{true}}
\quad {\color{examplegray}(confidence 5)}
\\[1pt]

Fix type &
consent or policy enforcement; data exposure restriction
\\[1pt]

GDPR principles &
lawfulness, fairness and transparency; purpose limitation;
data minimisation; integrity and confidentiality
\\[1pt]

GDPR articles &
Art.~5, 6, 32
\\[1pt]

Fix location &
initial commit
\\[1pt]

Fixed commit &
\texttt{3d61dbfd}
\\[1pt]

Buggy commit &
before this PR
\end{tabularx}

\end{tcolorbox}

\caption{An abridged \dsname{} entry from \emph{Verified}. The entry
combines repository metadata, the privacy-relevant patch, and structured
LLM-generated annotations.}
\label{fig:example}
\vspace{-6mm}
\end{figure}

In this paper, we introduce \dsname{}, which closes three gaps in existing data.
(1)~\emph{Traceability.} Privacy-law mining studies classify issues and commits by topic or change type, but do not link a concrete privacy defect to its accepted fix and review~\cite{hennig2026whos,kapitsaki2025evolution,kapitsaki2026analyzing}. \dsname{} provides this traceability by pairing each privacy defect, such as personal data written to logs or a missing consent check, with the merged fix and its review discussion (Fig.~\ref{fig:example}).
(2)~\emph{Coverage.} Existing privacy benchmarks are limited in scope; for example, GDPR-Bench-Android covers violations from only 15 Android repositories~\cite{ran2025gdprbench}. \dsname{} contains 1,169 privacy fixes and 1,643 non-fix PRs from 1,048 repositories across platforms, to our knowledge the largest cross-project dataset of labeled privacy fixes, with each fix further labeled by fix type and GDPR Art.~5(1) principles.
(3)~\emph{Evaluation.} Existing code review benchmarks target functional defects~\cite{zeng2025swrbench,zhang2026ccrab} or security~\cite{zhao2026sere}, not privacy. \dsname{} therefore provides \emph{Verified}, a manually checked test split of 250 privacy fixes and 250 non-fix PRs, together with an evaluation harness that turns real fixes into review tasks for systematically comparing reviewers, prompts, and coding agents (Section~\ref{sec:evaluation}).

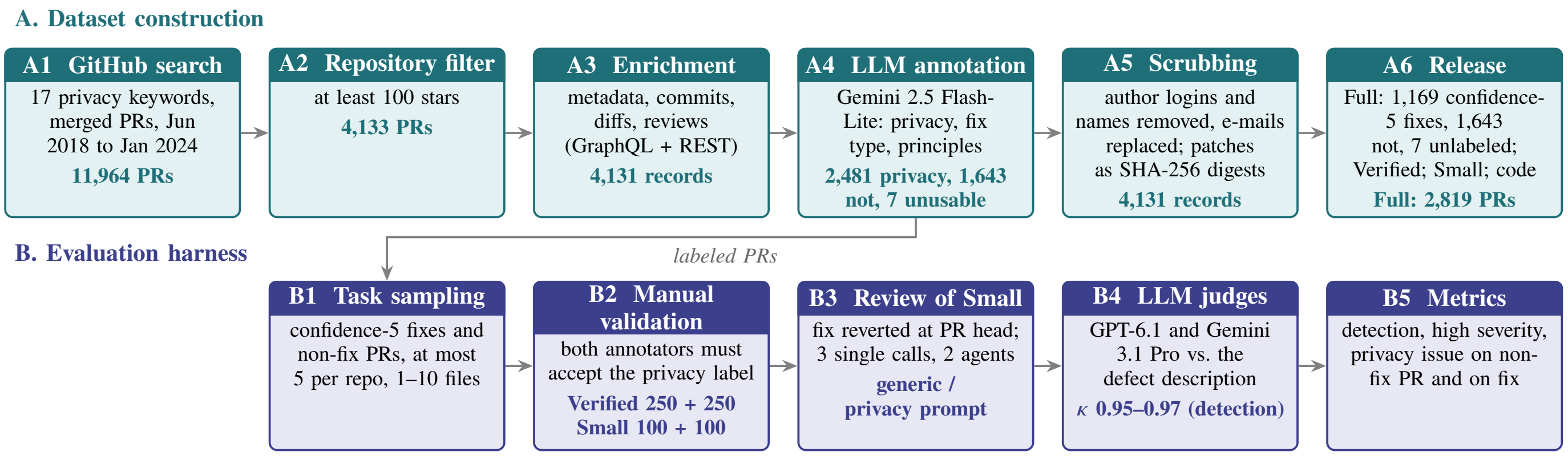
\begin{figure*}[ht]
\centering
\begin{tikzpicture}[
  arr/.style={-{Stealth[length=5pt,width=4pt]}, line width=0.7pt, draw=black!50},
  hd/.style={font=\footnotesize\bfseries}]
\newlength{\bw}\setlength{\bw}{0.150\textwidth}\newlength{\bh}\setlength{\bh}{1.95cm}\newlength{\hh}\setlength{\hh}{0.4cm}\newlength{\bs}\setlength{\bs}{0.169\textwidth}
\newcommand{\pbox}[7]{%
  \node[draw=#4dark, fill=#4fill, rounded corners=2.5pt, line width=0.7pt,
    minimum width=\bw, minimum height=\bh, inner sep=0, anchor=north west] (#1) at (#2,#3) {};
  \begin{scope}
    \clip[rounded corners=2.5pt] (#1.north west) rectangle (#1.south east);
    \fill[#4dark] (#1.north west) rectangle ([yshift=-\hh]#1.north east);
  \end{scope}
  \node[anchor=center, text width=\dimexpr\bw-4pt\relax, align=center, font=\footnotesize\bfseries, text=white]
    at ([yshift=-0.5\hh]#1.north) {#5};
  \node[anchor=north, text width=\dimexpr\bw-7pt\relax, align=center, font=\scriptsize, inner sep=0]
    at ([yshift=-\dimexpr\hh+2pt\relax]#1.north) {#6\\[2pt]{\color{#4dark}\bfseries #7}};
}
\pbox{a1}{0}{0}{teal}{A1\ \ GitHub search}{17 privacy keywords, merged PRs, Jun 2018 to Jan 2024}{11,964 PRs}
\pbox{a2}{\bs}{0}{teal}{A2\ \ Repository filter}{at least 100 stars}{4,133 PRs}
\pbox{a3}{2\bs}{0}{teal}{A3\ \ Enrichment}{metadata, commits, diffs, reviews (GraphQL + REST)}{4,131 records}
\pbox{a4}{3\bs}{0}{teal}{A4\ \ LLM annotation}{Gemini 2.5 Flash-Lite: privacy, fix type, principles}{2,481 privacy, 1,643 not, 7 unusable}
\pbox{a5}{4\bs}{0}{teal}{A5\ \ Scrubbing}{author logins and names removed, e-mails replaced; patches as SHA-256 digests}{4,131 records}
\pbox{a6}{5\bs}{0}{teal}{A6\ \ Release}{Full: 1,169 confidence-5 fixes, 1,643 not, 7 unlabeled; Verified; Small; code}{Full: 2,819 PRs}
\foreach \i/\j in {a1/a2,a2/a3,a3/a4,a4/a5,a5/a6} \draw[arr] (\i.east) -- (\j.west);
\node[hd, color=tealdark, anchor=south west] at ([yshift=0.12cm]a1.north west) {A. Dataset construction};
\newlength{\by}\setlength{\by}{-2.7cm}
\pbox{b1}{\bs}{\by}{indigo}{B1\ \ Task sampling}{confidence-5 fixes and non-fix PRs, at most 5 per repo, 1--10 files}{}
{\setlength{\hh}{0.62cm}\pbox{b2}{2\bs}{\by}{indigo}{B2\ \ Manual\\validation}{both annotators must accept the privacy label}{Verified 250 + 250\\Small 100 + 100}}
\pbox{b3}{3\bs}{\by}{indigo}{B3\ \ Review of Small}{fix reverted at PR head; 3 single calls, 2 agents}{generic / privacy prompt}
\pbox{b4}{4\bs}{\by}{indigo}{B4\ \ LLM judges}{GPT-6.1 and Gemini 3.1 Pro vs.\ the defect description}{$\kappa$ 0.95--0.97 (detection)}
\pbox{b5}{5\bs}{\by}{indigo}{B5\ \ Metrics}{detection, high severity, privacy issue on non-fix PR and on fix}{}
\foreach \i/\j in {b1/b2,b2/b3,b3/b4,b4/b5} \draw[arr] (\i.east) -- (\j.west);
\node[hd, color=indigodark, anchor=south west] at ([yshift=0.1cm]a1.west |- b1.north west) {B. Evaluation harness};
\draw[arr] (a4.south) -- ++(0,-0.22cm) -| (b1.north)
  node[pos=0.18, below, font=\scriptsize\itshape, color=black!60] {labeled PRs};
\end{tikzpicture}
\caption{Pipeline of \dsname{}: (A) dataset construction; (B) the
evaluation harness, which builds the Verified and Small splits and reviews
Small (Section~\ref{sec:evaluation}).}
\label{fig:pipeline}
\vspace{-3mm}
\end{figure*}

\section{Dataset Construction}
\label{sec:method}

\subsection{Search and Filtering (Steps A1--A2)}
\label{sec:retrieval}

Fig.~\ref{fig:pipeline}A shows the construction pipeline. Step~A1 queries
GitHub's search API for merged, non-draft PRs whose title or body contains
one of 17 keywords (matched by word stem). Eleven come from privacy-law
mining studies~\cite{franke2024gdpr,kapitsaki2025evolution,kapitsaki2026analyzing,hennig2026whos}
(e.g., \emph{GDPR}, \emph{CCPA}, \emph{anonymization}, \emph{cookie banner},
\emph{privacy violation}) and six recurred in privacy PRs seen during
collection (e.g., \emph{user consent}, \emph{PII leak}, \emph{Right to
Erasure}). A pilot of about ten LLM-labeled PRs per term dropped seven terms
with at most 36\% privacy PRs. As the
API returns at most 1,000 results per query, we query each keyword one year
at a time and stop at 2,000 PRs so that frequent keywords do not dominate.
Four keywords hit this cap, and three of them stop early: \emph{GDPR} after
2019, \emph{cookie banner} after 2021 and \emph{user consent} after 2022.
Deduplicating by URL gives 11,964 PRs. Step~A2 keeps repositories with at
least 100 stars (as of March 2026), a proxy for active projects, leaving
4,133 PRs from 1,469 repositories. Keyword search is noisy (63\% false
positives in~\cite{hennig2026whos}), so annotation (Step~A4) separates
privacy fixes from other PRs.

\subsection{Enrichment and Storage (Step A3)}
\label{sec:enrichment}

In Step~A3, one GraphQL query per PR retrieves its metadata (title, body, author,
timestamps, branches) and up to 100 each of its commits, changed files,
closing issues, comments, reviews and review threads (50 comments per
thread); the diff of every changed file comes from the paginated REST
endpoint \texttt{/pulls/\{number\}/files}. Two PRs that no longer resolve
are dropped, leaving 4,131 records from the same 1,469 repositories.
Each record is one JSON object with all of the above, the patches (digests
in the release, Step~A5) and the LLM annotation with the raw response.

\subsection{Annotation (Step A4)}
\label{sec:annotation}

\begin{table}[t]
\centering
\caption{Annotation schema returned for every PR.}
\label{tab:schema}
\scriptsize
\renewcommand{\arraystretch}{0.95}
\setlength{\tabcolsep}{3pt}
\begin{tabularx}{\columnwidth}{@{}l>{\raggedright\arraybackslash}X@{}}
\toprule
Field & Values and meaning \\
\midrule
\texttt{is\_privacy\_related} & true if the PR's primary goal is to fix a privacy issue, e.g., faulty handling of personal data or missing privacy functionality \\
\texttt{fix\_type} & one or more of 8: data removal, data masking, access control, data exposure restriction, encryption or secure storage, logging fix, consent or policy enforcement, other \\
\texttt{gdpr\_principles} & one or more of the six principles of GDPR Art.~5(1) \\
\texttt{gdpr\_articles} & free list of GDPR articles, e.g., Art.~32 \\
\texttt{confidence\_level} & 1 (low) to 5 (high); tracks privacy relevance (Section~\ref{sec:quality}) \\
\texttt{reasoning}, \texttt{evidence} & free text justifying the label \\
\midrule
\multicolumn{2}{@{}l}{\emph{Auxiliary fields}} \\
\texttt{where\_is\_the\_fix} & \emph{initial commit}, \emph{during review} (a reviewer flagged the issue and a later commit fixed it), or \emph{cannot determine} \\
\texttt{fixed\_commit}, \texttt{buggy\_commit} & commit that fixed the defect; commit that introduced it (99\% \emph{before this PR}) \\
\bottomrule
\end{tabularx}
\end{table}

\textbf{LLM annotation.} In Step~A4, every PR is annotated by Gemini 2.5 Flash-Lite
(\texttt{gemini-2.5-flash-lite}), chosen for its low cost. The prompt renders the PR as text (title, description,
commits, changed files, closing issues, comments, reviews and threads, and
diffs truncated to 3,000 characters per file, which affects 21.3\% of
patches) and defines the JSON output of Table~\ref{tab:schema} with one
worked example and criteria for \texttt{is\_privacy\_related} (the PR
addresses the handling or protection of personal data or mitigates risks
such as data leakage). The auxiliary fields rest on commit messages, as the model sees no per-commit
diffs. A second parser recovers 132 of 139 unparseable raw responses; seven PRs
have no usable label.

\textbf{Manual validation} (Fig.~\ref{fig:pipeline}, B2). Two recruited annotators, each holding at least a master's degree and two years of software engineering experience, complete a training session on the GDPR principles and the annotation schema. They then independently check the 650 LLM-screened candidates of the test split (Section~\ref{sec:evaluation}), seeing the model's labels. 
Each judges \texttt{is\_privacy\_related} of every candidate as correct, partially correct or incorrect from the PR page and diff, and the other fields of the 300 fixes, except \texttt{buggy\_commit} and confidence, as endorsed or not. Because they see the labels, the endorsement rates in Table~\ref{tab:valid} may be inflated. One annotator later checks 50 random Small defect descriptions.

\subsection{Scrubbing and Release (Steps A5--A6)}
\label{sec:availability}

Step~A5 removes author logins and names and replaces e-mail addresses and
the names written next to them; \texttt{@}mentions, branch names and PR
URLs remain. Patches, as third-party code, become SHA-256 digests that a
script rehydrates and verifies. Step~A6 releases \dsname{} at \dsdoi{} as
JSON Lines in three splits: \emph{Full}, the 1,169 PRs labeled privacy
fixes with confidence~5, the 1,643 non-fix PRs and the seven unlabeled ones
(2,819 PRs from 1,048 repositories); and \emph{Verified} and \emph{Small},
250 + 250 and 100 + 100 PRs whose privacy label both annotators accepted
(Section~\ref{sec:evaluation}). The 1,312 fixes
with confidence~2--4 are released separately, together with the collection
code, the harness, the Small reviews and judge verdicts, and the
annotators' verdicts. Data are under CC BY 4.0, code under MIT.

\subsection{Statistics and Label Quality}
\label{sec:quality}

\begin{table}[t]
\centering
\caption{\emph{Full}: PRs per LLM-assigned fix type and principle
(multi-label, \% of 1,169 fixes; four PRs with out-of-schema types
omitted) and per class (\% of 2,812 labeled). Median changed lines and
files; \% with a review text; \% labeled fixed during review; mean
comments; median days to merge.}
\label{tab:labels}
\scriptsize
\setlength{\tabcolsep}{2.2pt}
\begin{tabular}{@{}lrrrrrrr@{}}
\toprule
 & PRs (\%) & Lines & Files & Rev.\% & Dur.\% & Com. & Days \\
\midrule
\multicolumn{8}{@{}l}{\emph{Fix type}} \\
Exposure restriction & 807 (69.0) & 106 & 5 & 57.9 & 5.0 & 10.3 & 2.2 \\
Consent/policy & 405 (34.6) & 87 & 4 & 57.5 & 4.2 & 8.8 & 2.0 \\
Access control & 242 (20.7) & 402 & 12 & 71.5 & 5.8 & 17.9 & 4.4 \\
Other & 195 (16.7) & 51 & 4 & 47.7 & 3.1 & 10.0 & 1.2 \\
Data removal & 74 (6.3) & 333 & 10 & 64.9 & 4.1 & 17.8 & 5.5 \\
Data masking & 67 (5.7) & 110 & 3 & 43.3 & 6.0 & 5.1 & 2.1 \\
Logging fix & 37 (3.2) & 85 & 5 & 51.4 & 2.7 & 8.6 & 2.0 \\
Encryption & 10 (0.9) & 795 & 16 & 60.0 & 0.0 & 33.1 & 13.9 \\
\midrule
\multicolumn{8}{@{}l}{\emph{GDPR Art.~5(1) principle}} \\
Integrity \& confidentiality & 945 (80.8) & 115 & 5 & 58.6 & 5.0 & 10.4 & 2.6 \\
Lawfulness, fairness, transp. & 599 (51.2) & 63 & 3 & 53.9 & 5.0 & 7.7 & 1.6 \\
Data minimisation & 422 (36.1) & 107 & 5 & 55.5 & 3.6 & 10.6 & 2.1 \\
Purpose limitation & 142 (12.1) & 73 & 4 & 50.0 & 4.9 & 9.4 & 1.9 \\
Storage limitation & 7 (0.6) & 31 & 3 & 71.4 & 0.0 & 8.9 & 20.1 \\
Accuracy & 1 (0.1) & 6,204 & 100 & 100.0 & 0.0 & 144.0 & 68.5 \\
\midrule
\multicolumn{8}{@{}l}{\emph{Class}} \\
All privacy fixes & 1,169 (41.6) & 88 & 4 & 56.2 & 4.4 & 9.3 & 2.1 \\
Non-fix PRs & 1,643 (58.4) & 31 & 2 & 38.9 & 0.5 & 4.1 & 1.1 \\
\bottomrule
\end{tabular}
\vspace{-3mm}
\end{table}

\begin{table}[t]
\centering
\caption{Human check of the 650 screened candidates: \% endorsed by
annotators A and B (privacy label: judged correct), raw agreement, Gwet's
AC1 and Cohen's $\kappa$; \texttt{gdpr\_articles} over the 72 fixes with one.}
\label{tab:valid}
\scriptsize
\setlength{\tabcolsep}{3pt}
\begin{tabular}{@{}lrrrrrr@{}}
\toprule
Field & PRs & A\,\% & B\,\% & Agr.\,\% & AC1 & $\kappa$ \\
\midrule
\texttt{is\_privacy\_related}, fix & 300 & 94.0 & 88.7 & 89.0 & .88 & .31 \\
\texttt{is\_privacy\_related}, non-fix & 350 & 96.3 & 81.4 & 81.7 & .80 & .14 \\
\texttt{fix\_type} & 300 & 92.3 & 51.0 & 56.0 & .26 & .10 \\
\texttt{gdpr\_principles} & 300 & 96.7 & 70.7 & 70.0 & .59 & .02 \\
\texttt{gdpr\_articles} & 72 & 91.7 & 69.4 & 72.2 & .60 & .18 \\
\texttt{fixed\_commit} & 300 & 93.3 & 84.0 & 80.7 & .76 & .06 \\
\texttt{where\_is\_the\_fix} & 300 & 98.0 & 69.0 & 68.3 & .56 & .00 \\
\bottomrule
\end{tabular}
\vspace{-3mm}
\end{table}

\textbf{Distribution} (Table~\ref{tab:labels}). Of the 4,124 PRs with a usable
label, 2,481 are labeled privacy-related, 1,169 with confidence~5. Non-fix
PRs mostly get confidence~1 (89.8\%), so the field reads as privacy
relevance; the fixes span the whole window (148--265 per year). Half of
the fixes (49.2\%) combine several types. Only 260 fixes (22.2\%) cite a
GDPR article, 91 of them exactly Art.~5 and 32, as in the prompt's
example, so these counts and the integrity share may echo it.

\textbf{Typical fixes.} Integrity and confidentiality, the most frequent
principle, mostly concerns personal data leaking through logs, debug output
or APIs: kubernetes \#108631 drops a user-identifying field from the HTTP
log, and smart-answers \#5720 filters vaccination status from the
application log.
Lawfulness, fairness and transparency fixes add or repair consent flows
and notices: wp-calypso \#34187 shows the GDPR banner at signup, and
the PR in Fig.~\ref{fig:example} blocks e-mails sent without consent. Data
minimisation fixes stop collecting or transmitting identifiers that are
not needed: fastlane \#12869 enables IP anonymization in analytics, and
kubermatic \#3854 drops user name and e-mail from its metrics.

\textbf{Code changes and review.} Privacy fixes are larger, more discussed
and slower to merge than non-fix PRs (Table~\ref{tab:labels}), most of all
access control, data removal and encryption fixes. The LLM labels 52 fixes
(4.4\%) as raised by a reviewer and fixed in a later commit, an upper bound
(see below).

\textbf{Label quality} (Table~\ref{tab:valid}). A and B are the two human
annotators of Section~\ref{sec:annotation}. On the privacy label they
agree on 82--89\% of the PRs but $\kappa$ is low: most verdicts are
\emph{correct}, which raises chance agreement, and they questioned mostly
different, borderline PRs. On the finer fields, both endorse 72\% of the principle labels and 52\%
of the fix types in Verified; the rest are mostly labels that A endorses
and B questions (both reject only 4 and 19 of 300), which keeps $\kappa$
near chance. Users should restrict to labels both endorsed, or inspect
the contested ones, with the released verdicts.
\texttt{fixed\_commit} often points at a clean-up or merge commit, and
few \emph{during review} labels are endorsed, so the 52 review-raised
fixes are an upper bound.

\textbf{Precision of Full.} A and B judge 299 and 298 of the 300 confidence-5
test fixes at least partly privacy fixes; as GPT-6.1 screened them and
larger PRs are not covered, the precision of \emph{Full} is likely lower.

\section{Use Case: Benchmarking Privacy-Aware Code Review}
\label{sec:evaluation}

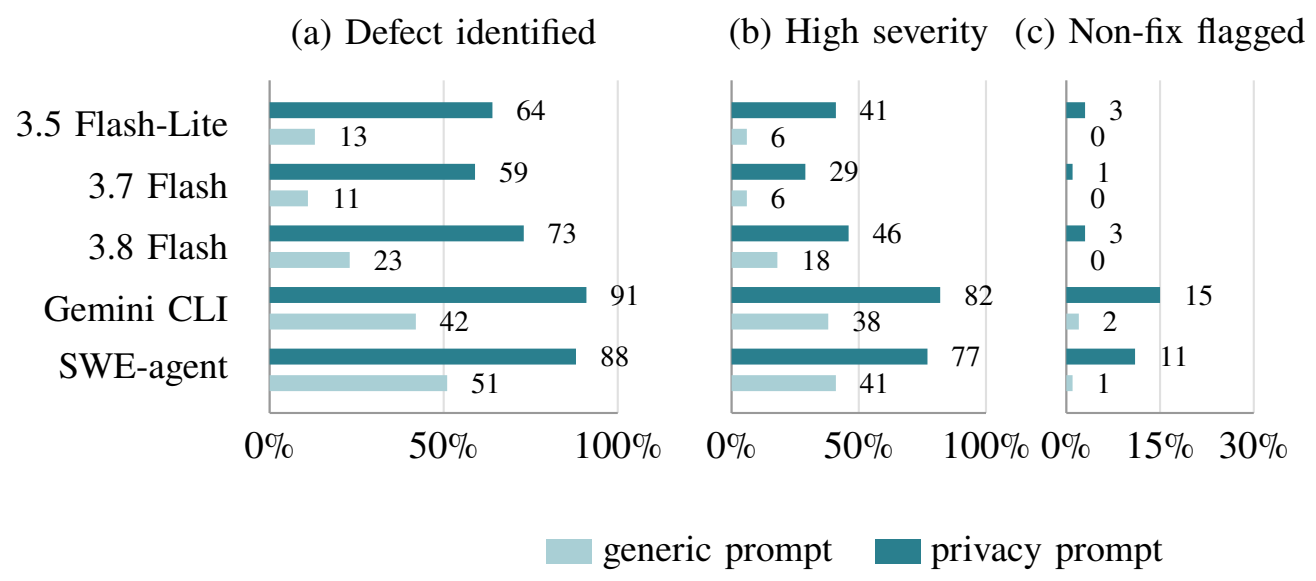
\begin{figure}[t]
\centering
\definecolor{barcol}{HTML}{2A7F8E}
\definecolor{barlight}{HTML}{A9CFD5}
\pgfplotsset{evalbar/.style={
  xbar, bar width=3pt, xmin=0, xmax=100, scale only axis, height=2.2cm,
  xtick={0,50,100}, xticklabel={\pgfmathprintnumber{\tick}\%},
  axis x line*=bottom, axis y line*=left, axis line style={black!40},
  xmajorgrids, grid style={black!12}, tick style={draw=none},
  y dir=reverse, ymin=-0.7, ymax=4.7, ytick={0,...,4},
  tick label style={font=\scriptsize}, title style={font=\scriptsize, yshift=-4pt},
  nodes near coords={\pgfmathprintnumber[fixed,precision=0]{\pgfplotspointmeta}},
  every node near coord/.append style={font=\tiny, text=black, anchor=west, xshift=1pt},
  clip=false}}
\begin{tikzpicture}
\begin{axis}[evalbar, name=pa, width=0.26\columnwidth,
  title={(a) Defect identified},
  yticklabels={3.5 Flash-Lite,3.7 Flash,3.8 Flash,Gemini CLI,SWE-agent}]
\addplot[fill=barlight, draw=none] coordinates {(13,0) (11,1) (23,2) (42,3) (51,4)};
\addplot[fill=barcol, draw=none] coordinates {(64,0) (59,1) (73,2) (91,3) (88,4)};
\end{axis}
\begin{axis}[evalbar, name=pb, width=0.19\columnwidth, at={(pa.south east)}, xshift=0.085\columnwidth, anchor=south west,
  title={(b) High severity}, yticklabels={,,,,}]
\addplot[fill=barlight, draw=none] coordinates {(6,0) (6,1) (18,2) (38,3) (41,4)};
\addplot[fill=barcol, draw=none] coordinates {(41,0) (29,1) (46,2) (82,3) (77,4)};
\end{axis}
\begin{axis}[evalbar, name=pc, width=0.14\columnwidth, at={(pb.south east)}, xshift=0.06\columnwidth, anchor=south west,
  title={(c) Non-fix flagged}, yticklabels={,,,,}, xmax=30, xtick={0,15,30},
  legend entries={generic prompt, privacy prompt},
  legend style={font=\scriptsize, draw=none, fill=none, at={(-1.1,-0.3)}, anchor=north, legend columns=2, /tikz/every even column/.append style={column sep=6pt}},
  legend image code/.code={\draw[#1, draw=none] (0cm,-0.08cm) rectangle (0.3cm,0.08cm);}]
\addplot[fill=barlight, draw=none] coordinates {(0,0) (0,1) (0,2) (2,3) (1,4)};
\addplot[fill=barcol, draw=none] coordinates {(3,0) (1,1) (3,2) (15,3) (11,4)};
\end{axis}
\end{tikzpicture}
\caption{Review rates on Small (\%, GPT-6.1 judge): (a), (b) over 100
reverted fixes, (c) over 100 non-fix PRs; 6 Gemini CLI runs hit the turn limit.}
\label{tab:eval}
\vspace{-7mm}
\end{figure}

\textbf{Verified and Small.} The harness (Fig.~\ref{fig:pipeline}B) draws, in a
seeded order (B1), privacy fixes with confidence~5 and, as controls,
\emph{non-fix PRs} labeled not privacy-related, by descending confidence;
each class has at most five PRs per repository and 1--10 changed files, at
least one of them code. GPT-6.1 describes each candidate fix's defect from
its label and diff, or reports none, and checks from the diff alone that
each non-fix PR fixes no privacy defect; inconsistent candidates are
discarded. The first 250 remaining candidates per class whose label both
annotators judged correct (B2) form Verified; the first 100 per class form Small. Selection favors small,
clean PRs (673 of 1,169 confidence-5 fixes qualify) and easy controls.

\textbf{Reviewers} (B3). Three single calls, Gemini 3.5 Flash-Lite, 3.7 Flash
and 3.8 Flash, span three cost tiers and see only the changed files and the
diff. Two open-source coding agents, Gemini CLI 0.64.0 and SWE-agent 1.1.0,
run on 3.7 Flash, so that comparing them with a single 3.7 Flash call
isolates what the agent adds; they review inside the full repository
(Docker, 60-minute limit, no web access; all runs October 2026).

\textbf{Test design} (B3--B5). Each PR of Small is a task that checks out the repository at its head
and presents the reversed diff, without documentation and changelogs, as the
change to review, so for a fix it reintroduces the defect. Each fix is also
reviewed forward, to see whether its known defect is claimed, and the
non-fix PRs measure false alarms.
Each reviewer gets a \emph{generic} and a \emph{privacy} review prompt. Two
judges, GPT-6.1 (\texttt{gpt-6.1-sol}) and Gemini 3.1 Pro (preview), decide
whether a review identifies the described defect, raises a privacy issue on
a non-fix PR or claims that defect in the original fix (Cohen's $\kappa$
between them 0.70--0.97); we report
GPT-6.1's verdicts (Gemini 3.1 Pro's detection rates differ by at most 4
points).

\textbf{Results} (Fig.~\ref{tab:eval}). Asking about privacy raises
detection for every reviewer (exact McNemar test, $p < 0.001$); 95\%
bootstrap intervals over repositories span about $\pm$10 points. On the same
model, the agents identify 88--91\% of the defects, 77--82\% as high
severity, against 59\% and 29\% for a single call, but also raise privacy
issues on 11--15\% of the non-fix PRs, against 1\%. Some may be real, as
these PRs match privacy keywords. With the privacy prompt, the agents
also claim the known defect in 13--21\% of the original fixes (single
calls: 4--10\%), so they are more sensitive but less specific. Agents also cost
more: 3.6--5.2 minutes and \$0.14--0.18 per review vs 14~s and \$0.013
for one call.

\textbf{Takeaways.} First, the prompt matters
more than the model tier: the cheapest single call with the privacy prompt
(64\%) beats the best single call (23\%) and both agents (42--51\%)
under the generic prompt. Second, repository access is what the agents add:
on the same model, detection rises from 59\% to 88--91\%, but at about ten
times the cost and ten times the false alarms. Third, the agents claim a
defect in 13--21\% of the correct fixes, about as often as they flag
non-fix PRs, which suggests that they react to code that touches personal
data rather than reason about the defect. Unless asked, every tested
reviewer misses about half or more of these defects.

\section{Discussion}
\label{sec:discussion}

\textbf{Uses and implications.} Researchers can study how privacy defects
arise, are discussed and are fixed, and which fix types and principles
recur. Tool builders can test classifiers of privacy-relevant changes on
\emph{Verified}, train them on the rest of \emph{Full}, and compare
reviewers, prompts and models on \emph{Small}. Practitioners who adopt agent-based review should ask for
privacy explicitly (Section~\ref{sec:evaluation}); the fix types offer a
checklist for such prompts, and the principle labels link each fix to a GDPR obligation.

\textbf{Limitations.}
\dsname{} relies on keyword retrieval and a single LLM annotation run, so it may miss privacy fixes and retain label noise; human validation covered only relatively small PRs and annotators saw the model labels. The schema maps CCPA and cookie-rule fixes to GDPR principles and omits some concepts, such as accountability and data-subject rights. The evaluation also has construct-validity risks: GPT-6.1 generates defect descriptions, screens tasks, and judges reviews; reverting fixes may expose easier cues than naturally occurring defects; each agent setting was run once; and all evaluated reviewers use Gemini models that may have seen the public fixes during training.

\textbf{Ethics.} We process public GitHub data for research as a legitimate
interest (GDPR Art.~6(1)(f)), with the scrubbing of Step~A5 as an Art.~89(1)
safeguard; this paper and the README inform developers (Art.~14(5)(b)), who
can ask hyang45@ncsu.edu to be removed. PR text went to the
Vertex AI and OpenAI APIs, which do not train on API data.

\section{Related Work}
\label{sec:related}

Prior work mines privacy law in repositories: Hennig et al.\ coded 652
issues on data protection~\cite{hennig2026whos}, Franke et al.\ compared
15,731 ``GDPR'' PRs with others by activity metrics~\cite{franke2024gdpr},
and Kapitsaki et al.\ coded 594 commits and 1,186 issues mentioning privacy
laws by relevance, topic and whether they change code or
text~\cite{kapitsaki2025evolution,kapitsaki2026analyzing}; these works label
samples by hand or not at all. For compliance detection, GDPR-Bench-Android
has 1,951 manually annotated violations in 15 Android
repositories~\cite{ran2025gdprbench}, LLMs~\cite{alecci2025toward} and
policy checks~\cite{andow2020policheck} target apps, and static analysis
points reviewers to code that processes personal
data~\cite{tang2023helping}. Among code review benchmarks,
SWR-Bench~\cite{zeng2025swrbench} and c-CRAB~\cite{zhang2026ccrab} evaluate
review on PRs or diffs, and SeRe~\cite{zhao2026sere} and
SecureReviewer~\cite{liu2026securereviewer} target security, as do the
vulnerability-fix datasets CVEfixes~\cite{bhandari2021cvefixes} and
MoreFixes~\cite{akhoundali2024morefixes}; none labels privacy compliance.
Unlike these works, \dsname{} pairs defects with accepted fixes and review,
labels every PR, and ships a verified test split with a harness.

\section{Conclusion}
\label{sec:conclusion}

This paper presents \dsname{}, a cross-project dataset of 1,169 privacy
fixes and 1,643 non-fix PRs from 1,048 GitHub repositories, with LLM-generated
privacy labels and human-validated test splits. Our evaluation shows that
current reviewers, including coding agents, detect substantially more privacy
defects when explicitly prompted to review for privacy. 
We release all artifacts for reproducible research.

\bibliographystyle{IEEEtran}
\bibliography{ref}

\end{document}